\documentclass[10pt,letterpaper]{article}         %
\usepackage{opex3}                                %
\usepackage{graphicx}
\usepackage{epsfig}
\usepackage{cite}
\usepackage{color}
\usepackage{amsmath,amssymb,subfigure}
\def\U#1{{\rm #1}} 
\def\u#1{_{\rm #1}}

\newcommand{\dagg}[1]{#1 ^\dagger}

\def\Dv{\U{D}\u{V}}
\def\Dt{\U{D}\u{T}}
\def\BGv{\U{BG}\u{V}}
\def\BGt{\U{BG}\u{T}}

\begin{document}
\title{
Observation of two output light pulses 
from a partial wavelength converter 
preserving phase of an input light 
at a single-photon level 
}
\author{
Rikizo~Ikuta,$^{1,*}$
Toshiki~Kobayashi,$^{1}$
Hiroshi~Kato,$^{1}$\\
Shigehito~Miki,$^{2}$
Taro~Yamashita,$^{2}$
Hirotaka~Terai,$^{2}$
Mikio~Fujiwara,$^{3}$
Takashi~Yamamoto,$^{1}$
Masahide~Sasaki,$^{3}$
Zhen~Wang,$^{2}$
Masato~Koashi,$^{4}$ and 
Nobuyuki~Imoto$^{1}$}

\address{
$^1$Graduate School of Engineering Science, Osaka University,
Toyonaka, Osaka 560-8531, Japan\\
$^2$Advanced ICT Research Institute, 
National Institute of Information and Communications Technology (NICT),
Kobe 651-2492, Japan\\
$^3$Advanced ICT Research Institute, 
National Institute of Information and Communications Technology (NICT),
Koganei, Tokyo 184-8795, Japan\\
$^4$Photon Science Center, 
The University of Tokyo, Bunkyo-ku, Tokyo 113-8656, Japan
}

\email{$^*$ikuta@mp.es.osaka-u.ac.jp} 



\begin{abstract}
We experimentally demonstrate that 
both of the two output light pulses of different wavelengths 
from a wavelength converter 
with various branching ratios
preserve phase information of an input light 
at a single-photon level. 
In our experiment, 
we converted temporally-separated two coherent light pulses 
with average photon numbers of $\sim 0.1$ at 780 nm 
to light pulses at 1522 nm 
by using difference-frequency generation 
in a periodically-poled lithium niobate waveguide. 
We observed an interference 
between temporally-separated two modes 
for both the converted and the unconverted light pulses 
at various values of the conversion efficiency. 
We observed interference visibilities greater 
than $0.88$ without suppressing the background noises 
for any value of the conversion efficiency 
the wavelength converter achieves. 
At a conversion efficiency of $\sim 0.5$, 
the observed visibilities 
are 0.98 for the unconverted light and 0.99 for the converted light. 
Such a phase-preserving wavelength converter 
with high visibilities will be useful for 
manipulating quantum states encoded in the frequency degrees of freedom. 
\end{abstract}


\section{Introduction}
A photonic quantum interface for a wavelength conversion 
was proposed in 1990~\cite{Kumar1990}, 
and it has been widely studied 
for quantum-information applications such as 
an efficient up-conversion detector~\cite{Langrock2005, Rakher2010}, 
a frequency-domain quantum eraser~\cite{Takesue2008}, 
and wavelength conversion 
aiming at long-distance quantum communication 
based on quantum repeaters~\cite{Tanzilli2005, Dudin2010, Ikuta2011, Zaske2012, Ikuta2013,Ikuta2013-2}. 
The wavelength conversion between two optical fields 
through a nonlinear optical medium 
with a sufficiently strong pump light 
is described by an effective Hamiltonian 
of a beamsplitter~(BS) in a frequency domain~\cite{Raymer2010}. 
The wavelength conversion 
described by such a Hamiltonian is expected to 
have properties that 
phases of the converted and the unconverted light pulses  
take over from the phase of the input light, 
and that the conversion efficiency is tunable by changing the pump power. 
In some of the previous demonstrations, 
it was observed that the converted light 
preserved the phase of the input light 
by using a weak coherent light 
at a single-photon level~\cite{Takesue2010, Curtz2010} 
and 
by using entangled photons~\cite{Tanzilli2005, Dudin2010,Ikuta2011,Ramelow2012}. 
On the other hand, to our best knowledge, 
none of demonstrations observes that 
an unconverted light retains 
phase information of an input light 
except the work of Giorgi {\it et. al.}~\cite{Giorgi2003} 
in which they observed the phase preservation of 
converted and unconverted light pulses 
with interference visibilities of 
$\sim 0.5$ and $\sim 0.7$, respectively, 
at a conversion efficiency of $\sim 0.4$. 
When the conversion efficiency increases, 
the signal-to-noise ratio of the unconverted mode decreases 
because the strong pump light increases the optical noises 
through the conversion process~\cite{Zaske2011, Ma2012} 
while the light in the unconverted mode decreases. 
When the conversion efficiency approaches unity, 
the signal-to-noise ratio will be almost zero. 
Moreover, in reality the conversion efficiency saturates 
at a level below unity, which cannot be explained by the simple Hamiltonian. 
Therefore, 
it is not obvious that 
the unconverted light holds phase information of the input light 
for high conversion efficiencies, 
and it is worth observing the phase preservation of 
not only the converted light but also the unconverted light 
for evaluating the basic property of 
the wavelength conversion working in a quantum regime. 

In this paper, 
we demonstrate that 
two output light pulses of different wavelengths 
after wavelength conversion inherit 
phase information of the input light with high visibilities. 
Observed interference visibilities are over $0.88$ 
without suppressing the background noises 
regardless of the conversion efficiencies 
our wavelength converter achieves. 
At a conversion efficiency of $\sim 0.5$, 
the observed visibilities are 0.98 for the unconverted light 
and 0.99 for the converted light. 
The wavelength conversion is achieved 
by the difference-frequency generation~(DFG) 
in a periodically-poled lithium niobate~(PPLN) waveguide. 
We used temporally-separated two light pulses 
at a single-photon level from a laser source at 780 nm 
as the input. 
The wavelength of the photons is 
converted to 1522 nm with various conversion efficiencies 
by choosing the pump power at 1600 nm. 
We observed an interference between temporally-separated two modes 
for both the converted and the unconverted light pulses. 
The operation of the partial wavelength converter 
which splits an input light into two different wavelengths 
while preserving the phase information is similar to 
the conventional beamsplitter dividing the input into two spatial modes. 
Therefore such a wavelength converter will be useful 
for manipulating quantum states encoded in the frequency degrees of freedom. 

\section{Theory of an ideal wavelength conversion}
We summarize the quantum dynamics 
of an ideal wavelength conversion 
based on difference frequency generation~(DFG) 
in a nonlinear optical medium as follows~\cite{Kumar1990, Ikuta2011}. 
We suppose that a signal mode at angular frequency $\omega\u{s}$ 
and a converted mode at angular frequency $\omega\u{c}$ 
satisfies $\omega\u{c}=\omega\u{s} - \omega\u{p}$, 
where $\omega\u{p}$ is the angular frequency of the pump light. 
When the pump light is sufficiently strong, 
the effective Hamiltonian of the DFG process is described by 
\begin{eqnarray}
{\hat{H}}=i\hbar (\xi^*\dagg{\hat{a}}\u{c}\hat{a}\u{s}
-\xi\dagg{\hat{a}}\u{s}\hat{a}\u{c}), 
\label{eq:Hamiltonian}
\end{eqnarray}
where 
$\hat{a}\u{s}$ and $\hat{a}\u{c}$ are annihilation operators 
of the signal mode and the converted mode, respectively. 
Here 
$\xi = |\xi|e^{i\phi}$ is proportional to the complex amplitude 
of the classical pump light. 
By using Eq.~(\ref{eq:Hamiltonian}), 
annihilation operators $\hat{a}\u{s, out}$ and $\hat{a}\u{c, out}$ 
of the signal and converted modes from the nonlinear optical medium 
are described by 
\begin{eqnarray}
\hat{a}\u{c,out}&=
 e^{-i\phi}\sin(|\xi| \tau)\ \hat{a}\u{s} +\cos(|\xi| \tau)\ \hat{a}\u{c} 
\label{eq:ac}
\end{eqnarray}
and 
\begin{eqnarray}
\hat{a}\u{s,out}&=
\cos(|\xi| \tau)\ \hat{a}\u{s} - e^{i\phi}\sin(|\xi| \tau)\ \hat{a}\u{c},
\label{eq:as}
\end{eqnarray}
where $\tau$ is the propagation time of the pulses 
through the nonlinear optical medium. 
Eqs.~(\ref{eq:ac}) and (\ref{eq:as}) are equivalent to 
the relation between input and output modes of a BS. 
The transmittance and the reflectance 
are $|\cos(\xi\tau)|^2$ and $|\sin(\xi\tau)|^2$, respectively. 
These can be adjusted by changing 
the pump power for the wavelength conversion. 
From Eqs.~(\ref{eq:ac}) and (\ref{eq:as}), 
the converted light and the remaining unconverted light 
take over the phase information from the input signal light. 
In the experiments in~\cite{Tanzilli2005, Dudin2010,Takesue2010, Curtz2010, Ikuta2011, Ramelow2012}, 
the phase preservation of the converted light in Eq.~(\ref{eq:ac}) 
has been demonstrated. 
On the other hand, 
that of the unconverted light in Eq.~(\ref{eq:as}) 
has not been observed with a high fidelity~\cite{Giorgi2003}. 

\section{Experiment}
\subsection{Experimental setup}
\begin{figure}[t]
 \begin{center}
 \scalebox{0.5}{\includegraphics{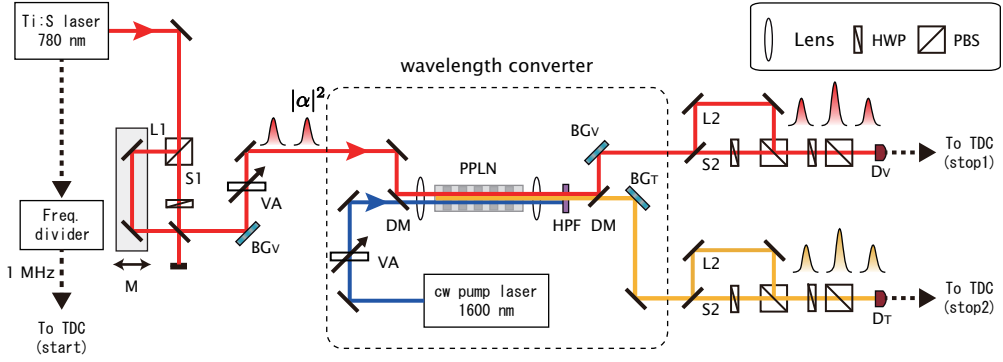}}
  \caption{ Experimental setup. 
  We initially prepare two temporally-separated light pulses at 780 nm. 
  Their frequencies are down-converted to 1522 nm 
  by DFG in the PPLN. 
  The length of the PPLN crystal is 20 mm, 
  and the acceptable bandwidth is calculated to be 0.3 nm. 
  The conversion efficiency is changed by the pump power 
  up to $\sim$ 700 mW. 
  After the process of the wavelength conversion, 
  the interference fringe of 
  each of the unconverted light at 780 nm and 
  the converted light at 1522 nm is measured. 
  \label{fig:setup}}
 \end{center}
\end{figure}
The experimental setup is shown in Fig.~\ref{fig:setup}. 
We use a $+45^\circ$ polarized 
mode-locked Ti:sapphire~(Ti:S) laser~(wavelength: 780 nm; 
pulse width: 1.2 ps; repetition rate: 82 MHz) as a light source. 
The light is divided into a short path~(S1) and a long path~(L1) 
by a polarizing beamsplitter~(PBS). 
After the polarization of the light passing through S1 is flipped 
from horizontal~(H) to vertical~(V) polarization 
by a half wave plate~(HWP), 
they are recombined by a BS. 
The time difference of about $600$ ps 
gives phase difference between the two light pulses, 
and it is varied by mirrors~(M) on a piezo motor driven stage. 
After the light pulses are spectrally narrowed 
by a Bragg grating~($\BGv$) with a bandwidth of 0.2 nm, 
an average photon number of each of the temporally-separated light pulses 
is set to $|\alpha|^2$, which can be varied from $10^{-3}$ to $1$ 
by a variable attenuator~(VA). 
Then the light pulses are coupled to 
the quasi-phase-matched PPLN waveguide~\cite{Nishikawa2009}. 
Their frequencies are down-converted to the wavelength of 1522 nm 
by the DFG using a cw pump laser at 1600 nm 
which is combined with the signal light by a dichroic mirror~(DM). 
The conversion efficiency is changed by the pump power 
up to $\sim$ 700 mW. 

After the wavelength conversion, 
the pump light is eliminated by a high-pass filter~(HPF), 
and the light pulses at 780 nm and 1522 nm are separated by a DM. 
The temporally-separated light pulses at 780 nm are 
diffracted by another $\BGv$, 
while the light pulses at 1522 nm are 
diffracted by $\BGt$ with a bandwidth of 1 nm. 
The temporally-separated two light pulses at each wavelength 
are divided into a short path~(S2) and a long path~(L2) 
with a time difference of $600$ ps. 
After the polarization of the light passing through S2 is flipped, 
the light pulses from S2 and L2 are mixed by a PBS. 
Finally, the light pulses are projected 
onto $+45^\circ$ polarization, 
and then coupled to single-mode fibers 
followed by 
superconducting single-photon detectors~(SSPDs)~\cite{Miki2009, Miki2010} 
which are denoted by $\Dv$ for the light at 780 nm 
and $\Dt$ for the light at 1522 nm. 

Electric signals from $\Dv$ and $\Dt$ are 
connected to a time-to-digital converter~(TDC) 
which is gated by a 1-MHz clock signal. 
The clock signal is obtained 
by frequency division of the 82 MHz clock signal from Ti:S laser. 
There are three possible arrival times 
in the electric signals from $\Dv$ and $\Dt$. 
The earlest and the latest signals are obtained 
by the light passing through S1-S2 and L1-L2 paths, 
respectively.  
We post-select the 200-ps time windows 
of the central peaks originated with the light pulses 
passing through S1-L2 and L1-S2. 
Such post-selected signals from $\Dv$ and $\Dt$ 
should show the first-order interference pattern 
of the coherent light pulses 
at 780 nm and 1522 nm, respectively, 
depending on the position of M. 

\subsection{Experimental results}
\begin{figure}[t]
 \begin{center}
 \scalebox{0.55}{\includegraphics{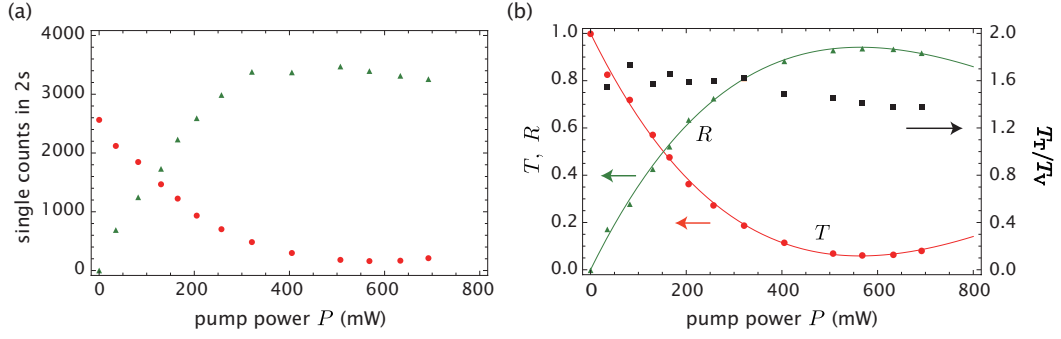}}
  \caption{
  (a) 
  Observed counts of the unconverted photons~(red circle) 
  and the converted photons~(green triangle). 
  (b) 
  The probability of the unconverted events $T$~(red circle), 
  the conversion efficiency $R=1-T$~(green triangle) 
  and the ratio of $T\u{T}$ to $T\u{V}$~(black square). 
  We derived the curve for $T(P)$ 
  by using the observed photon counts of the unconverted light. 
  The red curve for $T$ is obtained by the best fit 
  to $T$ with $1-A\sin^2(\sqrt{\eta P})$, 
  where $A\approx 0.94$ and $\eta \approx 0.0044/\U{mW}$. 
  The green curve for $R$ is given by $A\sin^2(\sqrt{\eta P})$. 
  \label{fig:conv}}
 \end{center}
\end{figure}
Before we describe the demonstration 
of the phase-preservation property of our wavelength converter, 
we first observed the dependencies of the probability of 
the unconverted events and the conversion efficiency 
on the pump power by measuring 
the photon counts of 
the unconverted mode at 780 nm and the converted mode at 1522 nm. 
In this experiment, we set $|\alpha|^2$ to $\sim 0.1$. 
The experimental result is shown in Fig.~\ref{fig:conv}(a). 
From the result, the maximum conversion efficiency is 
achieved at the pump power of $\approx 560$ mW 
which is smaller than the maximum pump power of 700 mW we can supply. 
We roughly estimate 
the internal conversion efficiency in the PPLN crystal as follows. 
The transmittance of the optical circuit 
before the wavelength conversion 
including the coupling efficiency to the PPLN is represented by $T\u{in}$. 
The conversion efficiency and the probability of the unconverted events 
are represented by $R(P)$ and  $1-R(P)=T(P)$, respectively, 
where $P$ is the pump power. 
Note that $R(P)$ corresponds to $|\sin(\xi\tau)|^2$ 
in Eqs.~(\ref{eq:ac}) and (\ref{eq:as}). 
We denote overall transmittance of the optical circuit 
including the quantum efficiency of the detector 
after the conversion process by 
$T\u{V}$ for the unconverted light at 780 nm. 
The detection counts of 
the unconverted light pulse is described by 
$C=\mathcal{N}T\u{in} T(P) T\u{V}$, 
where $\mathcal{N}$ is the total number of the input photons. 
We assume that $T\u{in}$ and $T\u{V}$ 
take constant values regardless of the pump power. 
By using $T(0\,\U{mW})=1$ and the observed count of 
$C_0=\mathcal{N}T\u{in}T\u{V}$ at $P=0$ mW, 
we obtain the dependency of $C/C_0=T(P)$ on the pump power 
as shown in Fig.~\ref{fig:conv}(b). 
The best fit to $T(P)$ with $1-A\sin^2(\sqrt{\eta P})$ 
gives $A\approx 0.94$ and $\eta \approx 0.0044/\U{mW}$. 

When we denote overall transmittance after the conversion process 
by $T\u{T}$ for the converted light at 1522 nm, 
the detection counts of the converted light is given 
by $\mathcal{N}T\u{in} R(P) T\u{T}$. 
Because $\mathcal{N}T\u{in} R(P) T\u{T}/C_0 
=R(P)T\u{T}/T\u{V}$ and $R(P)=1-T(P)$, 
$T\u{T}/T\u{V}$ is estimated 
from the observed photon count of the unconverted light 
and that of the converted light at each pump power, 
which is shown in Fig.~\ref{fig:conv}(b). 
We see that $T\u{T}/T\u{V}$ takes an almost constant value of about $1.5$, 
which is in accord with the assumption of $T\u{V}$ being constant 
in our rough estimation. 

\begin{figure}[t]
 \begin{center}
 \scalebox{0.55}{\includegraphics{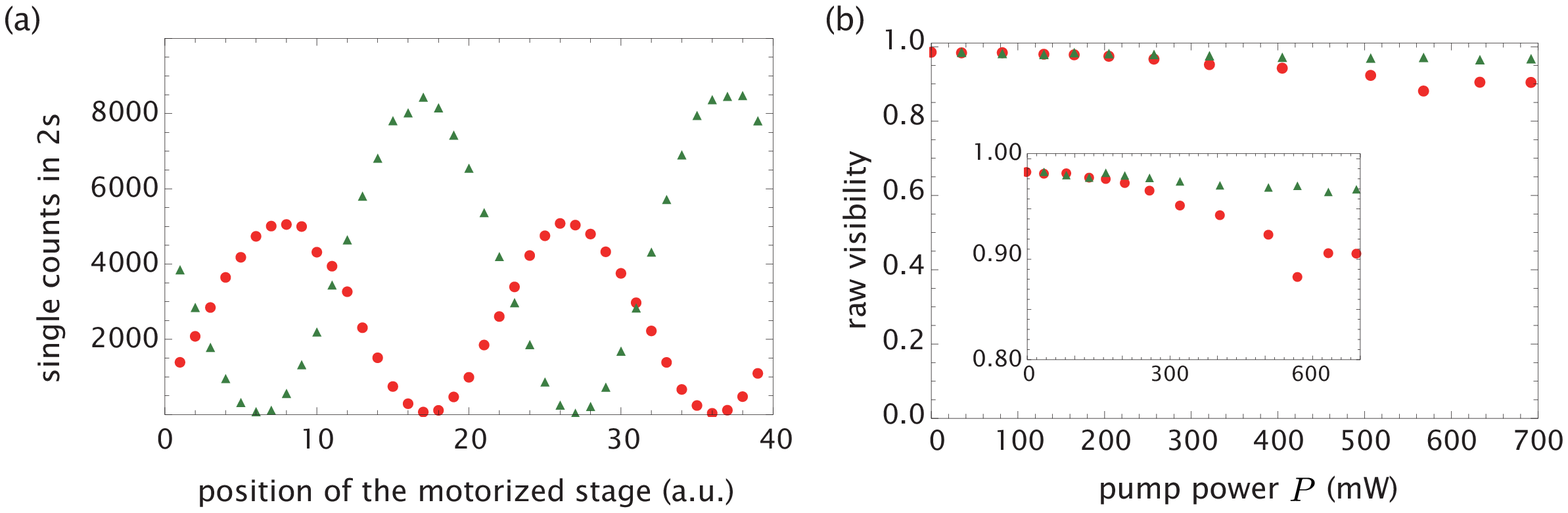}}
  \caption{
  (a)
  The interference fringes of 
  the unconverted photons~(red circle) 
  and the converted photons~(green triangle). 
  These data were observed at the pump power of 165 mW, 
  namely $T\approx 0.5$. 
  The interference visibilities are 0.98 for the unconverted light 
  and 0.99 for the converted light. 
  (b) Dependencies of the visibilities on the pump power. 
  The inset shows the enlarged view. 
  \label{fig:fringevis}}
 \end{center}
\end{figure}
Next we demonstrated that 
both the unconverted and the converted light pulses after 
the wavelength conversion preserve 
the phase information of the input light pulse. 
In this demonstration, the average photon number 
of the input light was set to $|\alpha|^2\approx 0.1$. 
By moving mirror M in Fig.~\ref{fig:setup}, 
the time difference 
between the light pulses passing through S1 and L1 is varied. 
As a result, the first-order interference fringe is observed 
after mixing the light pulses 
passing through the paths of S1-L2 and L1-S2. 
Fig.~\ref{fig:fringevis}(a) shows the experimental result 
of the interference fringe when the pump power is 165 mW. 
For both the unconverted and the converted light pulses, 
the interference fringes are clearly observed. 
We define the interference visibility by 
$V=(N\u{max}-N\u{min})/(N\u{max}+N\u{min})$, 
where $N\u{max}$ and $N\u{min}$ are 
the maximum and the minimum of the count rates, respectively. 
The observed values of the visibilities 
are $V=0.98$ for the unconverted light 
and $V=0.99$ for the converted light. 
The standard deviations of the visibilities 
with the assumption of the Poisson statistics of the counts 
are less than $0.01$. 
We then measured the visibilities of the interference 
for various values of the pump power 
ranging from $0\,\U{mW}$ to $700\,\U{mW}$. 
The experimental results are shown in Fig.~\ref{fig:fringevis}(b). 
For both the converted and the unconverted light pulses, 
we observed visibilities higher than $0.88$. 

\begin{figure}[t]
 \begin{center}
 \scalebox{0.55}{\includegraphics{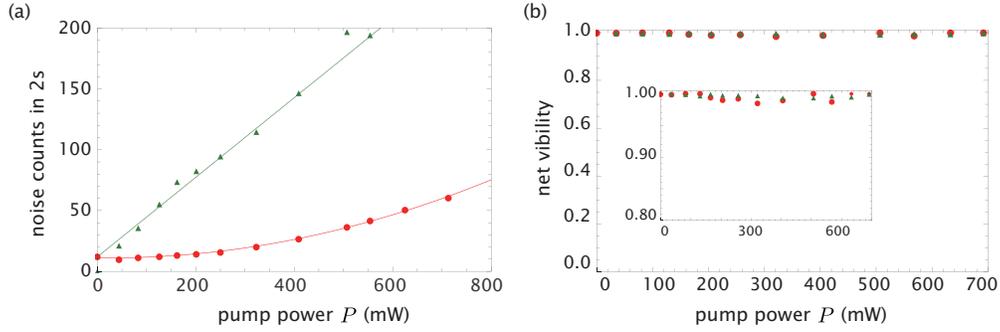}}
  \caption{
  (a) Dependencies of the background noises on the pump power. 
  The green triangles show the result for the unconverted visible photons. 
  The red circles show the result for the converted telecom photons. 
  The background noise of the unconverted mode 
  when the pump light is off is slightly larger than zero 
  while the dark count of the detector $\Dv$ is almost zero, 
  indicating that such a noise may be caused by 
  a component of the residual fundamental cw light 
  as a result of the imperfection of the signal light from the Ti:S laser. 
  (b) Dependencies of the net visibilities on the pump power. 
  The green triangles and the red circles show the results 
  for the unconverted visible and the converted telecom photons, respectively. 
  \label{fig:noise}}
 \end{center}
\end{figure}
From the experimental result in Fig.~\ref{fig:fringevis}(b), 
we see that the visibilities 
decrease with the pump power more prominently for the unconverted mode. 
In order to see the reason for the degradation of the visibilities, 
we measured the background noises 
from the detected counts 
temporally away from the three signal peaks. 
The average noise counts are shown in Fig.~\ref{fig:noise}(a). 
For the converted mode at 1522 nm, 
the linear dependency of the noise photons on the pump power is observed, 
which is caused by the Raman scattering 
of the pump light for the DFG~\cite{Ikuta2011, Zaske2011}. 
On the other hand, for the unconverted mode at 780 nm, 
nonlinear dependency of the noise photons on the pump power is observed. 
Such a dependency has been reported 
in a lot of demonstrations of frequency up-conversion 
to a visible light~\cite{Langrock2005, Pelc2011, Ma2012}. 
They claim that when we use a pump light 
longer than a signal light and a converted light, 
the optical noises are mainly originated 
from the Raman scattering of the pump light 
followed by its frequency up-conversion. 
By subtracting the background noises, 
we plotted the net visibilities in Fig.~\ref{fig:noise}(b). 
The net visibilities exceed 0.98 for all pump powers. 
The result indicates that the degradation of the visibilities 
are mainly caused by the optical noises from the pump light. 

\begin{figure}[t]
 \begin{center}
 \scalebox{0.47}{\includegraphics{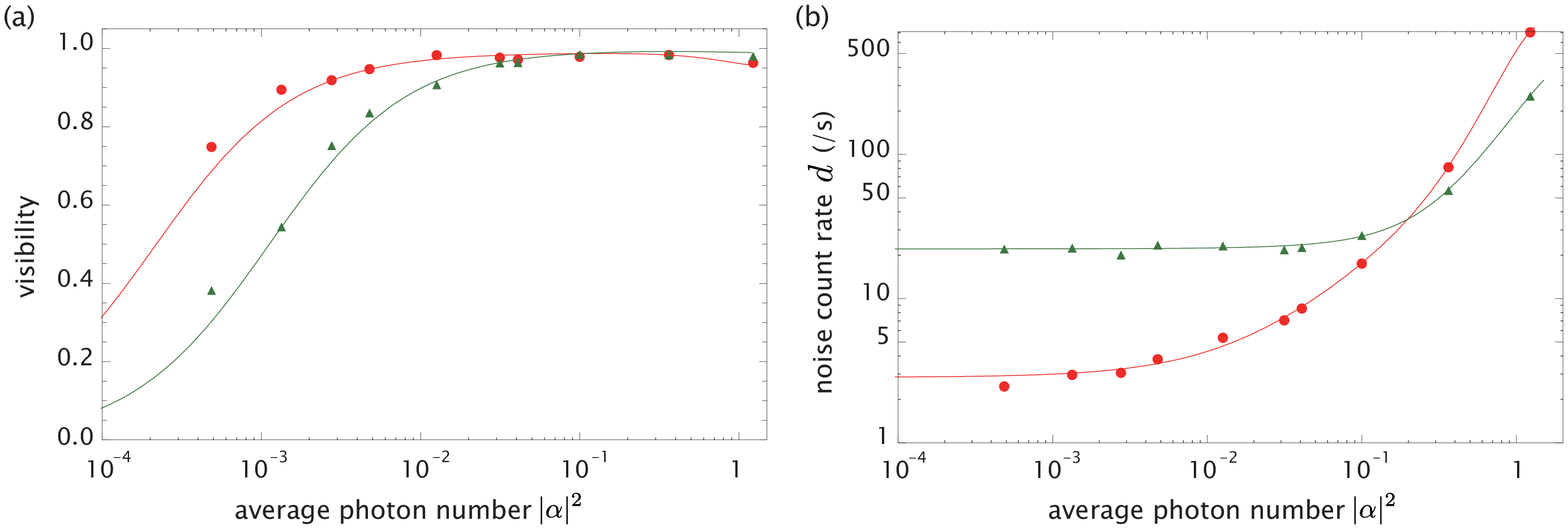}}
  \caption{
  (a) Dependencies of the visibilities on $|\alpha|^2$ 
  when the pump power is 165 mW. 
  The points of the red circle and the green triangle 
  show the visibilities of the unconverted photons 
  and the converted photons, respectively. 
  The solid curves are obtained by using the signal-photon rates 
  estimated from the experimental data and 
  the observed noise counts shown in Fig.~\ref{fig:alpha}(b). 
  (b) Dependencies of the noise counts on $|\alpha|^2$. 
  The points of the red circle and the green triangle 
  are for the unconverted photons 
  and the converted photons, respectively. 
  The solid curves are obtained 
  by using polynomial functions 
  fitted to the observed noise count rate $d$. 
  \label{fig:alpha}}
 \end{center}
\end{figure}
Finally, we measured the interference visibility 
for various values of $|\alpha|^2$ with the fixed pump power of 165 mW. 
The observed values of $V$ are shown in Fig.~\ref{fig:alpha}(a). 
The high visibilities over $0.9$ remain 
for the converted and the unconverted light pulses 
for $|\alpha|^2 > 0.01$. 
The behavior of the visibilities in Fig.~\ref{fig:alpha}(a) 
can be explained by using temporally continuous background noises 
depending on $|\alpha|^2$, 
which was separately measured and is shown in Fig.~\ref{fig:alpha}(b). 
We assume the noise photons are 
statistically independent of the signal photon counts. 
Due to the estimated values of the visibilities close to 1 
when we subtracted the background noises 
from the experimental result in Fig.~\ref{fig:noise}(b), 
we construct a simple model of the visibilities 
described by the signal photons with unit visibility and the noise photons. 
In this model, the visibility is given by 
$V=|\alpha|^2 T\u{all}f/(|\alpha|^2 T\u{all}f + 2 d)$, 
where $T\u{all}$ is the overall transmittance of the optical circuit 
described by $T\u{in} T(P) T\u{V}$ for the unconverted mode 
and by 
$T\u{in} R(P) T\u{T}$ for the converted mode. 
We roughly estimate 
$T\u{in}T\u{V}\approx 0.03$ for the unconverted mode 
and $T\u{in}T\u{T}\approx 0.04$ for the converted mode 
from the observed values. 
From Fig.~\ref{fig:conv}(b), we see 
$R(165\,\U{mW})=T(165\,\U{mW})=0.5$. 
$f=1$ MHz is the frequency of the clock 
and $d$ is the rate of the noise photons. 
By using polynomial functions 
fitted to the experimental result of 
the rates of the noise photons for $d$ 
as shown in Fig.~\ref{fig:alpha}(b), 
we obtain the expected curves of the visibilities 
shown in Fig.~\ref{fig:alpha}(a), 
which are in good agreement with the experimental data. 
From the high visibilities 
for $|\alpha|^2$ much smaller than $1$, 
the phase-preserving property of the wavelength converter 
will be expected to hold in a quantum regime. 
We note that 
the noise-photon rates take almost constant values for $|\alpha|^2 < 0.01$. 
These values are the intrinsic optical noises generated through the DFG. 
On the other hand, for larger values of $|\alpha|^2$, 
the noise-count rates increase. 
We guess this increase may come from 
the imperfection in the signal light from the Ti:S laser. 
Because the residual fundamental cw component of the laser 
is frequency-converted continuously, 
the effect of the cw component contributes 
to the constant background photon counts. 

\section{Conclusion}
In conclusion, we have demonstrated that the two output pulses 
at different wavelengths from the DFG-based wavelength converter 
using the PPLN crystal 
keep the phase information of the input light with high visibilities. 
By using the temporally-separated two coherent light pulses at 780 nm 
with average photon numbers smaller than $1$ 
as the input light pulse, 
we observed the interference 
between temporally-separated two modes 
for both the converted and the unconverted light pulses 
after the wavelength conversion. 
The observed values of the visibility are over $0.88$ 
for all conversion efficiencies achievable with our wavelength converter. 
At a conversion efficiency of $\sim 0.5$, 
the observed visibilities 
are 0.98 for the unconverted light and 0.99 for the converted light. 

\section*{Acknowledgments}
This work was supported by the Funding Program 
for World-Leading Innovative R \& D 
on Science and Technology~(FIRST), 
MEXT Grant-in-Aid for Scientific Research 
on Innovative Areas 21102008, 
MEXT Grant-in-Aid for Young scientists(A) 23684035, 
JSPS Grant-in-Aid for Scientific Research(A) 25247068 and (B) 25286077. 

\end{document}